\documentclass[aps,prl,twocolumn,superscriptaddress,groupedaddress]{revtex4}  
\pdfoutput=1

\usepackage{graphicx}  
\usepackage{dcolumn} 
\usepackage{bm}        
\usepackage{amssymb}

\begin{document}

\title{Crossed crystal scheme for fs-pulsed entangled photon generation in ppKTP}

\date{\today}
\author{Thomas Scheidl}
\email{thomas.scheidl@univie.ac.at}
\affiliation{Institute for Quantum Optics and Quantum Information, Austrian Academy of Sciences, Boltzmanng. 3, 1090 Vienna, Austria}
\author{Felix Tiefenbacher}
\affiliation{Institute for Quantum Optics and Quantum Information, Austrian Academy of Sciences, Boltzmanng. 3, 1090 Vienna, Austria}
\author{Robert Prevedel}
\affiliation{Faculty of Physics, University of Vienna, Boltzmanngasse 5, A-1090 Vienna, Austria}
\affiliation{Present address: Research Institute of Molecular Pathology (IMP) and Max F. Perutz Laboratories GmbH, Dr.-Bohr-Gasse 7-9, 1030 Vienna, Austria}
\author{Fabian Steinlechner}
\affiliation{ICFO--Institut de Ciencies Fotoniques, 08860 Castelldefels (Barcelona), Spain}
\author{Rupert Ursin}
\affiliation{Institute for Quantum Optics and Quantum Information, Austrian Academy of Sciences, Boltzmanng. 3, 1090 Vienna, Austria}
\affiliation{Vienna Center for Quantum Science and Technology, Faculty of Physics, University of Vienna, Boltzmanngasse 5, A-1090 Vienna, Austria}
\author{Anton Zeilinger}
\affiliation{Institute for Quantum Optics and Quantum Information, Austrian Academy of Sciences, Boltzmanng. 3, 1090 Vienna, Austria}
\affiliation{Vienna Center for Quantum Science and Technology, Faculty of Physics, University of Vienna, Boltzmanngasse 5, A-1090 Vienna, Austria}

\begin{abstract}
We demonstrate a novel scheme for femto-second pulsed spontaneous parametric down-conversion in periodically poled KTP crystals. Our scheme is based on a crossed crystal configuration with collinear quasi-phase-matching. The non-degenerate photon pairs are split in a fiber-based wavelength division multiplexer. The source is easier to align than common pulsed sources based on bulk BBO crystals and exhibits high-quality polarization entanglement as well as non-classical interference capabilities. Hence, we expect this source to be a well-suited candidate for multi-photon state generation e.g. for linear optical quantum computation and quantum communication networks. 
\end{abstract}

\maketitle

Photonic entanglement plays an important role in testing fundamental aspects of physics \cite{Schrodinger35a,Einstein35a} and serves as a key resource in modern quantum information processing (QIP) \cite{kimble}. The current gold standard for creating entanglement utilizes spontaneous parametric down-conversion (SPDC) in non-linear crystals \cite{Bunham70,Kwiat95,pelton04a}. For SPDC, one generally has to distinguish between continuous-wave (CW) and pulsed operation. In quantum communication experiments such as quantum key distribution, where multiple pair generation degrades the efficiency of the cryptographic protocols, CW pump lasers are the systems of choice. Multi-photon experiments, however, require the simultaneous generation of multiple pairs which are then combined to a multi-partite entangled state, e.g. a GHZ-state \cite{GHZ90}, a cluster-state, or a graph state \cite{briegel09a}, representing key ingredients for linear optical quantum computing and multi-party quantum communication networks. Engineering a multi-partite entangled state in this manner relies on non-classical interference between independent photons. In this respect, pulsed pump lasers can provide the timing required for coherently combining photons from independent pairs on linear optical devices such as (polarizing) beam splitters.

To date, the most efficient multi-photon sources employ femto-second ultra-violet pump lasers for generating near infra-red (NIR) photon pairs. Typically, these sources utilize consecutively aligned type-II phase matched BBO crystals \cite{huang11a,yao12a,yin12a,Ma12a}, such that independent  pairs can be generated in each of the crystals by the same pump pulse. Due to the particular phase matching conditions in BBO, photon pairs are emitted along the well-known SPDC cones, causing only the small fraction of pairs, generated in the overlap region of the two cones, to be entangled in their polarization degree of freedom. Spatial mode separation in these schemes is achieved by coupling the photons of a pair in two different optical single-mode fibers (SMF), which affects the stability and flexibility of experimental setups involving multi-photon states.

Periodic poling of non-linear crystals in combination with quasi-phase matching now offers ways to develop simpler and potentially more efficient schemes for pulsed entangled photon sources. Periodically poled crystals are best employed in a non-critical collinear quasi-phase matching configuration, in which spatial mode separation can be accomplished by dichroic or polarization splitting of the generated photon pair.  First attempts to employ periodically poled crystals for pulsed SPDC were undertaken in Refs. \cite{bannaszek01a,shi04a}, where non-entangled photon pairs in periodically poled KTP (ppKTP) were generated.  High-quality pulsed entangled photon pairs employing ppKTP in a Sagnac configuration have been produced in Refs. \cite{kuzucu07a,predojevic12a} using pico-second pump pulses. The reported spectral brightness (i.e. the number of entangled pairs/(s$\cdot$mW$\cdot$nm)) of both sources was significantly enhanced compared to standard pulsed BBO schemes, indicating the potential of employing such systems in multi-photon state generation. Non-classical interference between photons generated in separate periodically poled crystal has been investigated in the telecom regime using pico-second pumped ppLN waveguides \cite{aboussouan10}. However, so far it has not been reported in the most efficient regime for multi-photon state generation, where NIR photons are generated by femto-second pump pulses.

In this paper we present a new, femto-second pulsed SPDC source of polarization entangled NIR photons, based on ideas developed in Refs. \cite{kwiat99a,pelton04a}. We exploit collinear phase matching in ppKTP and show high quality entanglement of the two-photon state generated. Furthermore, the ultra-short pump pulses in combination with short ppKTP crystals and appropriate spectral filtering of the SPDC photons allow for non-classical interference, which we test in a Hong-Ou-Mandel-type interference experiment. Collinear emission of non-degenerate photon pairs enable us to couple both photons of a pair into a single optical SMF, in which spatial mode separation is subsequently accomplished by a fiber-based wavelength division multiplexer (WDM). Such a configuration significantly reduces the complexity of the alignment procedure and improves the flexibility of the whole setup as compared to standard BBO realizations. Due to its simplicity and brightness, our scheme is an excellent candidate for the use in multi-photon experiments.

\begin{figure}[t!]
                \centering
  \includegraphics[width=0.40\textwidth]{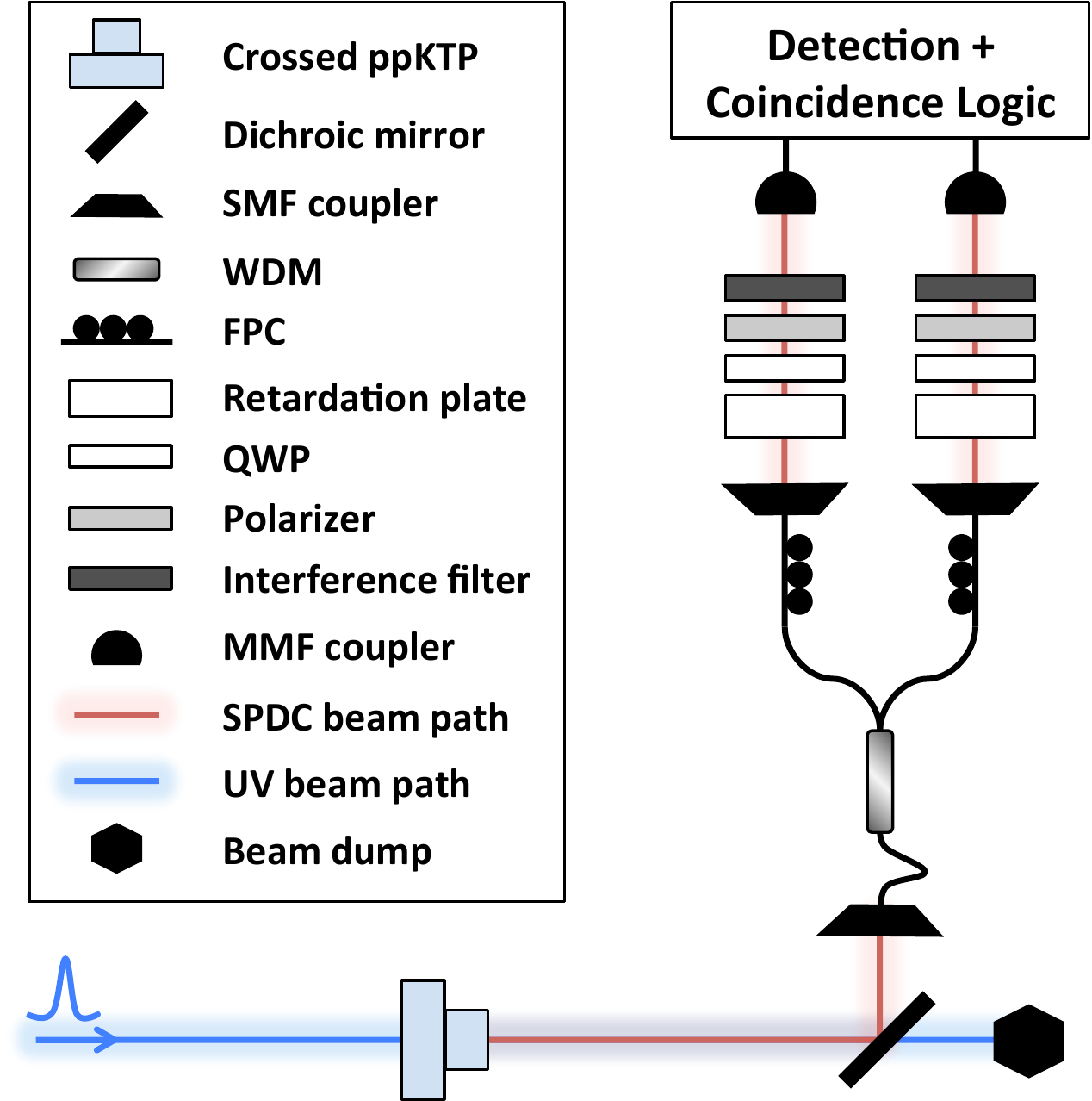}
                  \caption{(Color online) The setup of the fs-pulsed crossed-crystal source of entangled photons. Single-mode fiber coupler, SMF coupler; wavelength division multiplexer, WDM; fiber polarization controller, FPC; quarter-wave plate, QWP; multi-mode fiber coupler, MMF coupler. For details refer to the main text.}
                \label{setup}
\end{figure}

The setup of our pulsed entangled photon source is depicted in Figure~\ref{setup}. It utilizes two 1-mm-long X-cut ppKTP crystals placed in sequence, with their crystallographic Y~(Z)-axis oriented horizontally (vertically) and vertically (horizontally), in respective order. They are manufactured with a poling period of 7.9~$\mu$m for type-II collinear phase matching (Y$_p$Y$_s$Z$_i$) with pump (p), signal (s) and idler (i) photons having wavelengths of $\lambda_p=391.2$~nm, $\lambda_s=760$~nm and $\lambda_i=810$~nm, respectively. The pump pulses are generated by a frequency-doubled mode-locked Ti:Sapphire laser, polarized at 45$^\circ$ with respect to the crystallographic Y-axis, with a pulse duration of 150~fs and a repetition rate of 76~MHz.
The linearly polarized pump pulse can produce a pair of down-converted photons in either of the two crystals with equal probability. The first crystal generates photon pairs with the signal horizontally (H) and the idler vertically (V) polarized, whereas the second crystal does the same for polarizations interchanged. Since both creation processes occur with equal probability, the maximally entangled state
\begin{equation}
                \centering
  \psi=\frac{1}{\sqrt{2}}(|H\rangle_{s}|V\rangle_{i}-e^{-i\phi}|V\rangle_{s}|H\rangle_{i})
                \label{psi}
\end{equation}
is produced, as long as no additional information reveals in which crystal the pair was created. In order to erase any spatial "which-crystal" information, the non-degenerate signal and idler photons are, after being separated from the pump using a dichroic mirror, coupled into one and the same SMF and split with a fiber-based WDM into the respective output ports. The temporal "which-crystal" information, which arises due to the different group velocities of pump, signal and idler photons in the ppKTP crystal, is erased by guiding both photons of a pair through separate birefringent retardation plates of appropriate thicknesses (i.e. $\sim$3~mm of calcite for the signal photons and $\sim$3.6~mm for the idler photons). The retardation plates, together with additional polarization optics (i.e. quarter-wave plates and polarizers) and interference filters, are assembled in free-space. This allows investigating the polarization correlations and adjusting the spectral bandwidth of the generated entangled pairs. Additionally, the phase $\phi$ of  state (\ref{psi}) can be aligned by slightly tilting the retardation plates. Finally, the photons are coupled into multi-mode fibers (MMF) and detected using fiber-coupled silicon avalanche photo diodes, whereby two-fold detection events were identified using a coincidence counter with a coincidence time window of 4.4~ns. In a recent work, a qualitatively similar setup has been used to generate pulsed entangled photon pairs, however, using pico-second pump pulses and exploiting the process of four-wave mixing in cross-spliced birefringent fibers \cite{meyer-scott13}

As a first step, we evaluated the spectral brightness of our photon-pair source, utilizing interference filters with a full-width at half maximum (FWHM) bandwidth of 3~nm for both photons. For this purpose, the polarizers were removed from the setup.  With a pump power of 30~mW, we detected approximately 15000~coincidences/s with an overall coupling efficiency (coincidence-to-singles ratio) of 15\%. This corresponds to a spectral brightness of 170~coincidences/(s$\cdot$mW$\cdot$nm) and is already comparable to the pair generation rate of state-of-the-art entangled photon sources used for multi-photon state generation, where values between 30 and 240~coincidences/(s$\cdot$mW$\cdot$nm) have been reported \cite{huang11a,yao12a,yin12a,Ma12a}. The spectral brightness of our source could be further increased by approximately one order of magnitude using high quality interference filters ($>$90\% peak transmission) and more efficient single-photon detectors ($>$60\% detection efficiency) as well as optimized focusing of pump (the ideal pump beam-waist inside the crystal is $<10\mu m$) and collection optics \cite{bennink10a}. In comparison, our results were obtained employing filters with 75\% peak transmission, single-photon detectors with approximately 50\% detection efficiency, a pump waist of $\approx60\mu m$ and non-perfect collection of the photons from the crystals to the SMF. Note that, due to beam distortion, possibly stemming from photo-refractive effects \cite{taya96a} induced by the pump laser, the spectral brightness of our source decreased at average pump powers above 100~mW. This, however, does not pose a fundamental limitation to our scheme and can be mitigated using periodically poled crystals with increased resistance to photo-refractive damage.

\begin{figure}[t!]
                \centering
  \includegraphics[width=0.40\textwidth]{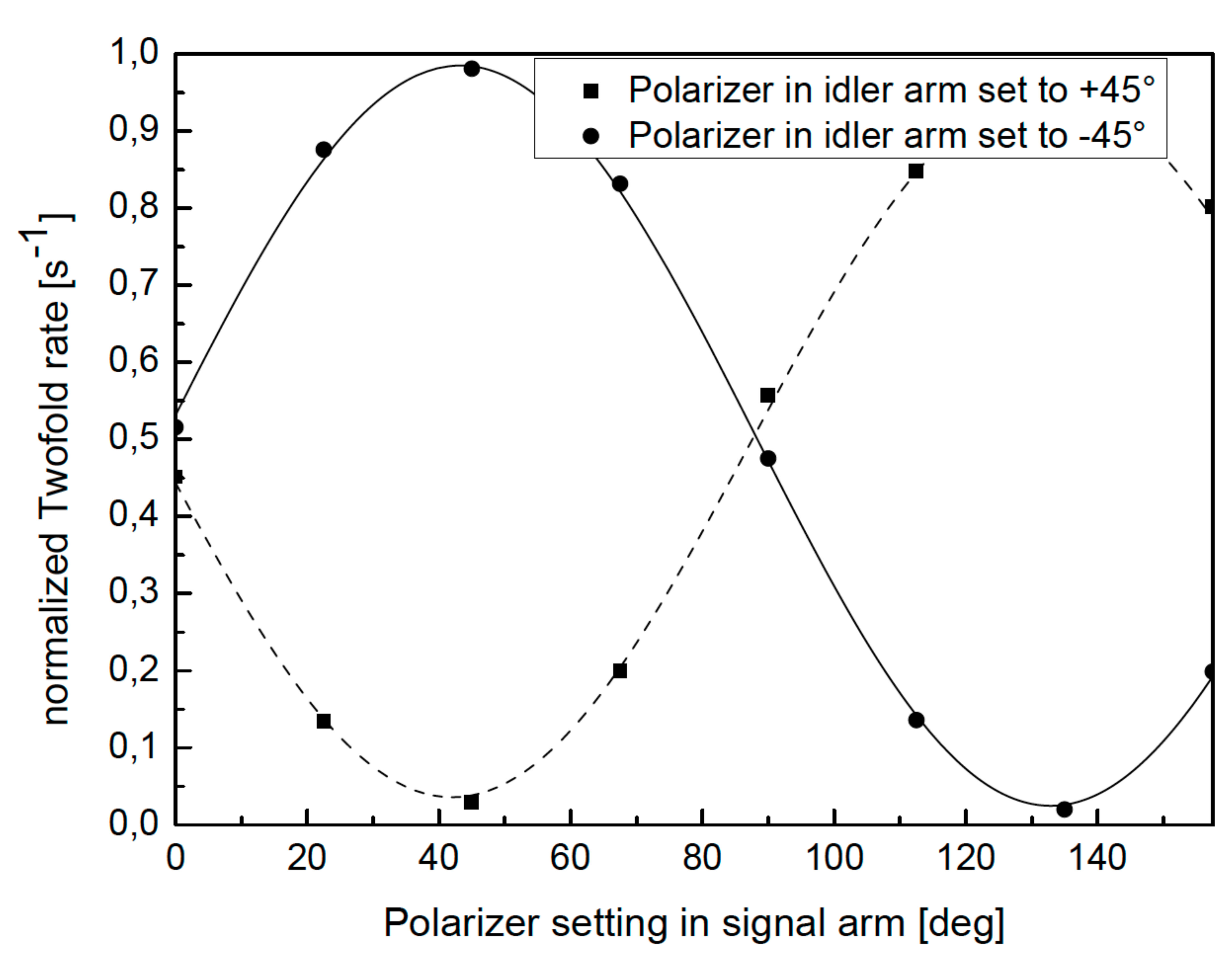}
                \caption{Measured polarization correlations of the generated entangled state in the $|\pm45^\circ\rangle$-basis, with $|\pm45^\circ\rangle=1/\sqrt{2}(|H\rangle\pm|V\rangle)$.  The polarizer in the idler arm was set to +45$^\circ$ (--45$^\circ$), respectively while the angle of the polarizer in the signal arm was rotated in steps of 22.5$^\circ$. The measured data points were fitted with a $sin^2$-function to guide the eye. The fringe visibility of the measured polarization correlations was calculated to be 96.0$\pm$0.2\%. The measurement duration was 10~s per datapoint -- errors, calculated using Poissonian photon statistics, are smaller than the size of the data points.}
                \label{bell}
\end{figure}

In order to characterize the quality of two-photon polarization entanglement of our source in a next step, we set the phase $\phi$ of  state (\ref{psi}) to generate the maximally entangled Bell state 
\begin{equation}
                \centering
  \psi^-=\frac{1}{\sqrt{2}}(|H\rangle_{s}|V\rangle_{i}-|V\rangle_{s}|H\rangle_{i}),
                \label{psiminus}
\end{equation}
by tilting one of the calcite retardation plates. A simple method for investigating the quality of the entangled state (\ref{psiminus}) is to measure the polarization correlations between the signal and idler photons in the $|\pm45^\circ\rangle$-basis, with $|\pm45^\circ\rangle=1/\sqrt{2}(|H\rangle\pm|V\rangle)$. The result of such a polarization correlation measurement at a pump power of 30~mW is shown in Figure \ref{bell} and yielded a fringe visibility of $V=96.0\pm$0.2\%. Furthermore, we used the generated two-photon state to test for a CHSH-type Bell inequality \cite{Bell64a,Clauser69a}. Such a Bell test requires 16 correlation measurements to reveal the Bell value $S$, and a measured value greater than 2 is a direct indication of the quantum nature of the generated state. Our measurements revealed $S=2.712\pm0.007$, which is in good agreement with the expected experimental value $S_{exp}=2\cdot\sqrt{2}\cdot V$, thus violating the inequality with more than 100 standard deviations statistical significance (errors were calculated using Poissonian photon statistics).

In order to further characterize the entangled state, we performed full quantum state tomography \cite{james01a} at several pump intensities. The fidelity $F=\langle\psi^-|\rho|\psi^-\rangle$ of the experimentally obtained two-photon density matrix $\rho$ with the maximally entangled state (\ref{psiminus}) was measured to be 95.8$\pm$0.2\%, 95.0$\pm$0.2\% and 92.6$\pm$0.2\% for pump intensities of 30~mW, 140~mW and 325~mW, respectively. These results compare quite well with the theoretical results in Refs. \cite{kuzucu07a,takesue10a}, indicating that the entanglement visibility in first order linearly decreases with the pump power due to higher order emissions. Hence, we infer that the quality of the entangled state is not influenced by the degradation of the ppKTP crystals at higher pump intensities, although the spectral brightness was negatively affected, as described above. The statistical standard deviations of these results were estimated by performing a 100 run Monte Carlo simulation of the state tomography analysis, with Poissonian noise added to the count statistics in each run \cite{james01a}. 

Finally, we used our two ppKTP crystals to study the feasibility of combining the produced photon pairs to a multi-partite state. As mentioned earlier, the generation of a multi-photon state relies on non-classical interference between independent single photons (i.e. photons from separate pairs), which can be investigated in a Hong-Ou-Mandel (HOM) type interference experiment \cite{hong87a}. The basic principle of such an experiment is to combine two photons from separate SPDC events on a 50/50 beam splitter. Given their indistinguishability in all degrees of freedom, the photons will bunch and leave the beamsplitter together through the same output port. However, photons created by independent SPDC events intrinsically carry a relative timing jitter due to the finite temporal width of the pump pulse and the group velocity mismatch (GVM) between the pump and SPDC photons in the crystal. This leads to temporal distinguishability at the beam splitter and degrades the interference  visibility. Nevertheless, by proper spectral filtering, the coherence and, hence, the indistinguishability can be restored.
\begin{figure}[t!]
                \centering
\includegraphics[width=0.49\textwidth]{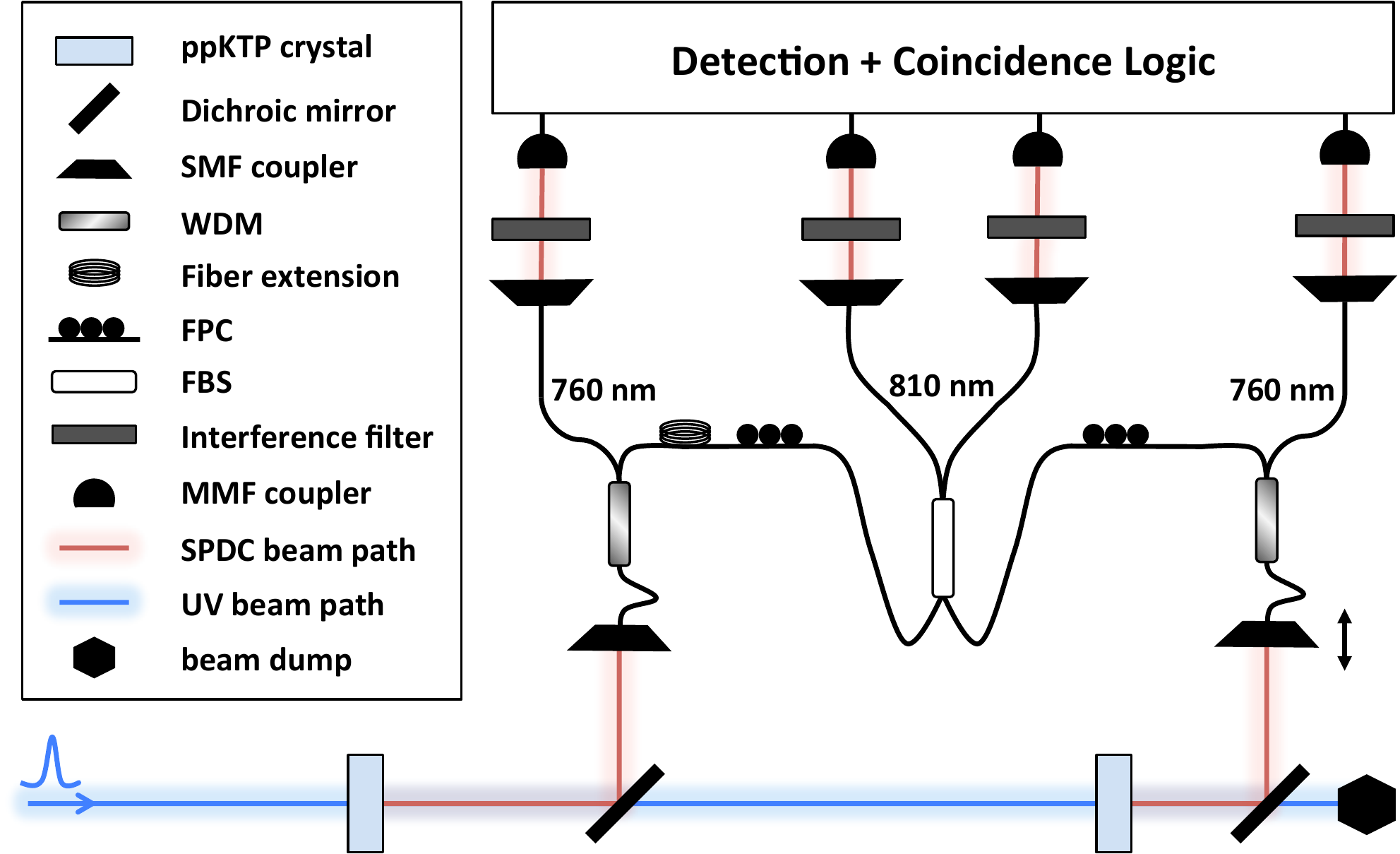}
                \caption{(Color online) An illustration of the setup used for investigating non-classical interference between photons generated in separate ppKTP crystals. Single-mode fiber coupler, SMF coupler; wavelength division multiplexer, WDM; fiber polarization controller, FPC; fiber beam-splitter, FBS; multi-mode fiber coupler, MMF coupler. For details refer to the main text. }
                \label{hom}
\end{figure}
The setup we used for testing the HOM-type interference is depicted in Figure \ref{hom}. We separated our two ppKTP crystals from the crossed-crystal configuration and used them to individually produce two independent non-entangled photon pairs. Both crystals, oriented with their crystallographic Y-axis horizontally, were consecutively pumped with horizontally polarized light from our pulsed laser. The SPDC photons from each crystal were separately split from the pump with dichroic mirrors and coupled into a SMF. After spatial mode separation, accomplished with the fiber based WDMs, the idler photons from both sources with a wavelength of 810~nm were combined on a 50/50 fiber beam splitter (FBS), thereby already ensuring spatial indistinguishability. The input ports of the FBS were attached to fiber polarization controllers for the purpose of guaranteeing indistinguishability in the polarization degree of freedom of the input photons. Additionally, to make the interfering photons indistinguishable in time, one of the SMF couplers could be varied in its position along the beam direction using a translation stage, which was controlled by a stepper motor with a resolution better than 10~$\mu$m. Each output of the FBS was connected to a fiber coupler, guiding the photons through equal interference filters such that spectral indistinguishability is provided.  Finally, the idler photons were coupled to MMFs and guided to the detectors. Note that, since the GVM between the pump and idler photons in our 1-mm ppKTP crystal is approximately 850~fs, this leads, in combination with the 150~fs pump pulses, to a temporal jitter between the idler photons from separate events of  $\sim$1~ps. Hence, we employed interference filters with a spectral FWHM bandwidth of 1~nm to ensure that the coherence time of the interfering photons covers this arrival-time uncertainty at the FBS.

\begin{figure}[t!]
                \centering
\includegraphics[width=0.40\textwidth]{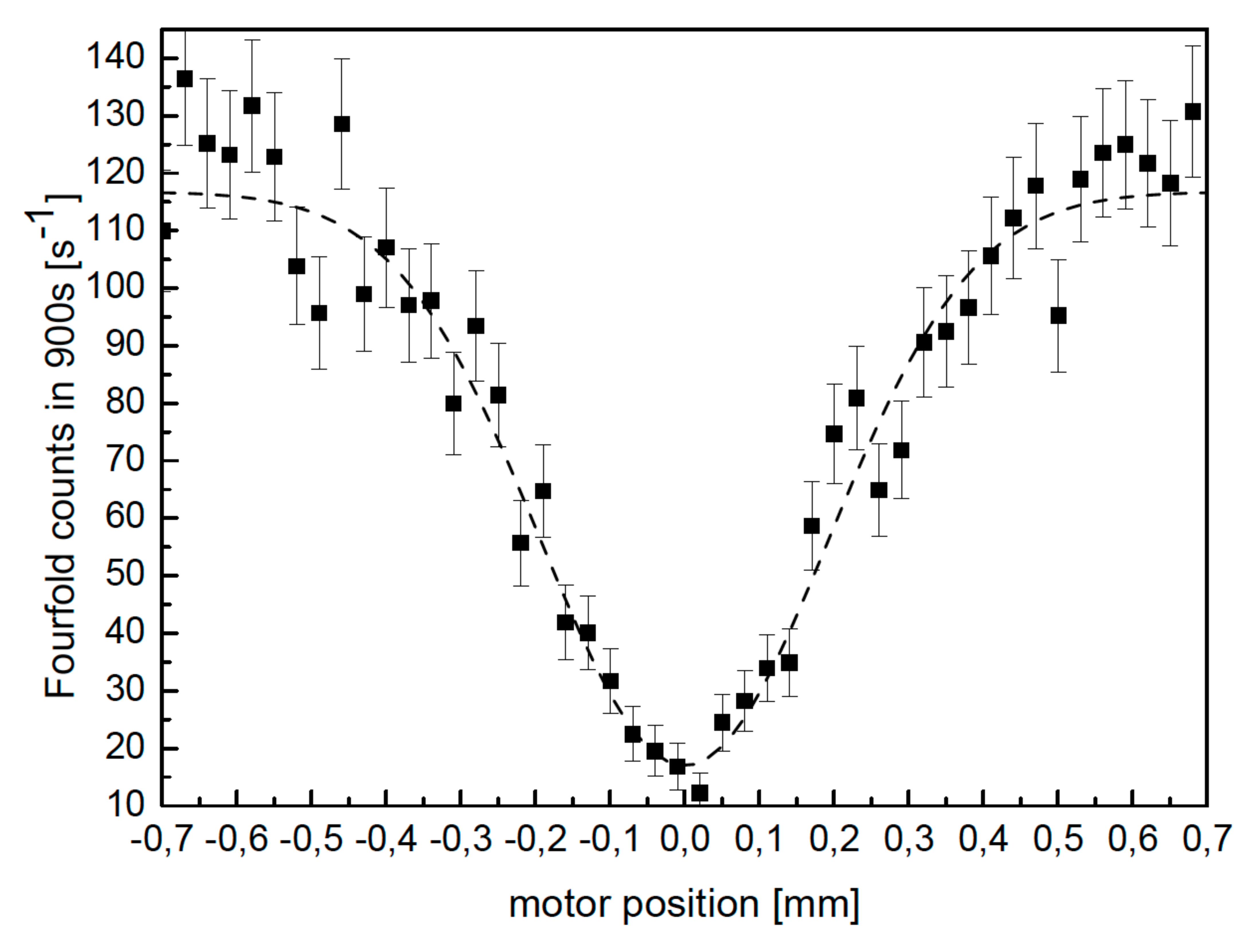}
                \caption{4-fold count rate in the HOM-type interference experiment as a function of the position of the motorized translation stage. When the photons coincide temporally, they overlap coherently at the beam splitter and photon bunching occurs, resulting in a dip in the 4-fold coincidence rate. The dip visibility of 88.8$\pm$0.1\% was calculated from the raw data, clearly showing the ability of the generated photons to interfere non-classically.}
                \label{hom_result}
\end{figure}

To prove that the observed interference is genuine, we utilized the signal photons (spectrally filtered to a FWHM bandwidth of 3~nm) from each pair as a trigger, thus heralding the presence of the corresponding idler photons. Therefore, we recorded 4-fold coincidences between the two output ports of the FBS and the two signal arms of the WDM. If the photons interfere coherently on the beam splitter, photon bunching occurs and both photons leave the beam splitter through the same output port. Hence, the 4-fold coincidence rate should ideally drop to zero. Varying the temporal delay between the idler photons around the optimal position using the motorized translation-stage, this drop in 4-fold rate can be scanned. In Figure \ref{hom_result},  the result of a typical measurement run is shown. The observed interference visibility $V$, calculated from the raw data without any background subtraction as $V=(I_{max}-I_{min})/I_{max}$, was 88.8$\pm$0.1\%, clearly proving the ability of our source for HOM-type interference. This is the first demonstration of high quality non-classical interference employing femto-second pulsed SPDC in separate periodically poled crystals.

In conclusion, we have presented a new scheme for femto-second pulsed SPDC to generate polarization entanglement, that is of considerable interest to multi-photon experiments. Our new scheme exploits non-degenerate collinear quasi-phase matching in ppKTP, such that spatial mode splitting can be accomplished with a fiber-based WDM. This facilitates the alignment and increases the flexibility of the setup, both crucial properties for multi-photon experiments. The obtained brightness is comparable to state-of-the-art sources used in multi-photon experiments and can further be enhanced by at least one order of magnitude using high-quality equipment and optimized focusing optics. The entanglement quality was verified via state tomography and a test of the CHSH inequality. Finally, the capability of non-classical interference was demonstrated in a HOM-type interference experiment between photons generated in separate ppKTP crystals, proving the feasibility of assembling a multi-photon entangled state out of individually generated photon pairs.

We acknowledge financial support by the FWF, the European Commission under the Integrated Project Qubit Applications (QAP) and the U.S. Army Research Office funded DTO (QCCM).

\end{document}